# Beyond quasi-optics: an exact approach to self-diffraction, reflection and finite-waist focusing of matter wave trajectories

Adriano Orefice[1], Raffaele Giovanelli[1] and Domenico Ditto[1]

*[1]University of Milan, Department of Agricultural and Environmental Sciences (DiSAA)*
*Via G. Celoria 2, Milan, ITALY 20133*
*e-mail address: adriano.orefice@unimi.it*

The "main road" open by de Broglie's and Schrödinger's discovery of matter waves and of their eigen-functions branched off, as is well known, into different "sub-routes". The most widely accepted one is Standard Quantum Mechanics (SQM), interpreting the *time-dependent* Schrödinger equation as the basic evolution law of a *wave-packet* which represents the simultaneous probabilical permanence of a particle in its full set of eigenstates. Another "sub-route" is offered by Bohm's Mechanics, able to reproduce the same results of SQM, while interpreting the stream-lines of the probability current density as the "quantum trajectories" of the moving particles. Reminding that the so-called *quasi-optical approximation* represents a standard mathematical technique allowing a *ray-based* treatment of any kind of wave-like features, we present here an *exact* wave-mechanical "sub-route", based on the observation that the *time-independent* Schrödinger equation (as well as any other Helmholtz-like equation) may be treated, bypassing the quasi-optical approximation, in terms of a Hamiltonian set of rays mutually coupled by an energy-dependent function (which we call "Wave Potential") encoded in the very structure of any Helmholtz-like equation. These rays, reducing to the *classical point-particle* trajectories when the Wave Potential is neglected, lend themselves to be interpreted as the *exact wave-dynamical trajectories and motion laws of classical-looking point-particles* associated with the de Broglie-Schrödinger matter waves. The role of the Wave Potential, acting *perpendicularly* to the momentum of the moving particles, is to "pilot" them without any energy exchange: a property which isn't shared by the well-known "Quantum Potential" of the Bohmian theory, involving the entire spectrum of possible eigen-energies of a wave-packet. This property turns out to allow the numerical computation of the particle trajectories, which we perform and discuss here for particles moving (under the guiding rule of the Wave Potential) in many different force-fields, such as a constant external field and the fields due to a potential barrier, a potential step and a lens-like potential, respectively.

## I. INTRODUCTION

As is well expressed in Ref. [1], "*the knowledge of several routes and their connections is always helpful when traveling through the quantum territory*". Different "routes" may complement each other, indeed, in different regions of that territory. But from which common "main road" shall the routes branch off? We think that the universally accepted common ground (simply called here "*Wave Mechanics*") is given by:

1. de Broglie's seminal relation $\vec{p} = \hbar \vec{k}$ [2, 3], verified beyond any doubt by the Davisson-Germer experiments [4], and establishing, once and for all, the objective reality of matter waves and the wave-particle duality;

2. the *time-independent* Schrödinger equation [5, 6], bypassing (with its eigen-fuctions and eigen-values) the heuristic prescriptions of the "old" Quantum Mechanics, and

3. the *time-dependent* Schrödinger equation, open to a wide spread of interpretations and developments.

Before passing to exploit, in Sect.V, a "route" of our own (the "Wave Potential" route), grafted on the common "main road" of Wave Mechanics and allowing, *without any further assumption*, an exact, trajectory-based treatment of point-particle dynamics, we shall begin by a brief summary (in Sects. II-IV) both of this "main road" and of its most successful "routes" (Standard Quantum Mechanics and Bohmian Mechanics) in order to develop a common language allowing a comparison (in Sects.VI and VII) of aims, methods, interpretations and results.

## II. WAVE MECHANICS

We shall refer, in order to fix ideas, to the case of non-interacting particles of mass $m$ and total energy $E$, launched with an initial momentum $\vec{p}_0$ (with $p_0 = \sqrt{2\,m\,E}$) into a force field deriving from a time-independent potential field $V(\vec{r})$. The classical dynamical behavior of each particle is described, as is well known [7], by the time-independent Hamilton-Jacobi (HJ) equation

$$(\vec{\nabla} S)^2 = 2\,m\,[E - V(\vec{r})] \quad , \qquad (1)$$

where the basic property of the HJ function $S(\vec{r}, E)$ is that the particle momentum is given by

$$\vec{p} = \vec{\nabla} S(\vec{r}, E). \qquad (2)$$

In other words, the (time-independent) *classical* HJ surfaces $S(\vec{r},E)=const$ are perpendicular to the momentum of the moving particles, and *pilot* them along *stationary* trajectories, according to the laws of Classical Mechanics. Louis de Broglie, reflecting on the analogy between the Maupertuis and Fermat variational principles [2, 3, 7], was induced to associate each material particle with a suitable "matter wave" of the form

$$\psi = u(\vec{r},\omega)\, e^{-i\,\omega\,t} \equiv R(\vec{r},\omega)\, e^{\,i\,[\varphi(\vec{r},\omega)-\omega t]}\,, \qquad (3)$$

with real amplitude $R(\vec{r},\omega)$, real phase $\varphi(\vec{r},\omega)$ and

$$E = \hbar\omega\,, \qquad (4)$$

according to the basic Ansatz

$$\vec{p}/\hbar \equiv \vec{\nabla}S(\vec{r},E)/\hbar = \vec{k} \equiv \vec{\nabla}\varphi\,, \qquad (5)$$

an Ansatz viewing the HJ surfaces $S(\vec{r},E)=const$ as the phase-fronts of these matter waves, while maintaining the *piloting role* played in Classical Mechanics.

The successive step, due to Schrödinger [5, 6], may be very simply performed [8, 9] by viewing Classical Mechanics, represented here by Eq. (1), as the eikonal approximation of a suitable Helmholtz-like equation that is immediately obtained, starting from Eqs. (3)-(5), in the form

$$\nabla^2 u(\vec{r},E) + \frac{2m}{\hbar^2}\,[E-V(\vec{r})]\,u(\vec{r},E) = 0\,, \qquad (6)$$

which is the usual form of the *time-independent* Schrödinger equation, holding for matter waves associated with particles of mass $m$ moving in an external *stationary* potential $V(\vec{r})$. This equation admits, as is well-known [8, 9], a (discrete or continuous, according to the boundary conditions) set of energy eigen-values and ortho-normal eigen-modes, which we shall indicate respectively (referring for simplicity to the discrete case) by $E_n$ and $u_n(\vec{r})$. From Eqs. (3)-(6) we get both the *ordinary-looking* wave equation

$$\nabla^2\psi = \frac{2m}{E^2}\,[E-V(\vec{r})]\,\frac{\partial^2\psi}{\partial t^2}\,, \qquad (7)$$

describing the dispersive character of the de Broglie matter waves associated with particles of total energy $E$, and the relation

$$\nabla^2\psi - \frac{2m}{\hbar^2}\,V(\vec{r})\,\psi = -\frac{2m}{\hbar^2}\,E\,\psi \equiv -\frac{2mi}{\hbar}\,\frac{E}{\hbar\omega}\,\frac{\partial\psi}{\partial t} = -\frac{2mi}{\hbar}\,\frac{\partial\psi}{\partial t}$$

that is

$$i\,\hbar\,\frac{\partial\psi}{\partial t} = -\frac{\hbar^2}{2m}\,\nabla^2\psi + V(\vec{r})\,\psi\,, \qquad (8)$$

which is the usual form of the *time-dependent* Schrödinger equation. Any wave-like implication of Eq. (8) (which is not, in itself, a wave equation) is due to its connection with the *time-independent* Schrödinger equation (6), *from which it is obtained*. Eqs. (6) and (8) arise therefore from a combined de Broglie's and Schrödinger's extension of Classical Mechanics, and don't need to be assumed (as it's sometimes done) as postulates.

By defining both the *eigen-frequencies* $\omega_n = E_n / \hbar$ and the *eigen-functions*

$$\psi_n(\vec{r},t) = u_n(\vec{r})\, e^{-i\,\omega_n\,t} \equiv u_n(\vec{r})\, e^{-i\,E_n\,t\,/\,\hbar} \qquad (9)$$

it's a standard procedure to verify that any linear superposition (with arbitrary constant coefficients $c_n$) of the form

$$\psi(\vec{r},t) = \sum_n c_n\, \psi_n(\vec{r},t)\,, \qquad (10)$$

is a (deterministically evolving) solution of the *time-dependent* Schrödinger Eq. (8). Since Eq. (8) *is not a wave equation*, the composite function (10) cannot represent an individual wave, revealable by a single Davisson-Germer experiment: it represents, in principle, *a collection* of individual de Broglie's matter waves $\psi_n(\vec{r},t)$, each one satisfying the wave equations (6) and (7) for an appropriate energy value $E_n$. Such a composite function could provide for instance a *weighted average* taken over the eigen-functions $\psi_n(\vec{r},t)$, where the coefficients $c_n$ (in duly normalized form) would represent either a set of experimental results (in view of a statistical treatment) or an *ad hoc* mathematical assembling, in view of the construction of a particular "packet" of wave-trains.

## III. STANDARD QUANTUM MECHANICS

Renouncing - both because of the uncertainty principle and because of the energy-independence of Eq. (8) - to a classical-looking particle dynamics, Max Born proposed, for the function (10), a role [10] going much beyond that of a simple superposition, assuming it to represent the most complete possible description ("Born's Wave-Function") of the physical state of a particle whose energy is not determined, in the form of a simultaneous permanence (before observation) in its full set of eigenstates, according to the probabilities $\left|c_n\right|^2$. The *continuous and deterministic* evolution of the "wave-packet" $\psi(\vec{r},t)$ according to Eq. (8) was associated to the further assumption of a *discontinuous and probability-dominated* process, after interaction with a measuring apparatus, causing its "collapse" into a single eigen-state. Even though "*no generally accepted derivation has been given to date*" [11], this "Born Rule" led, together with the uncertainty principle, to Standard Quantum

Mechanics (SQM), an intrinsically probabilistic conception of physical reality which was widely accepted as the pillar of any further development of microphysics.

Any system of $N$ particles with position vectors $\vec{r}_1, ..., \vec{r}_N$ is assumed to be described by a *single*, $3N$-dimensional Schrödinger equation with a *single* Wave Function $\psi(\vec{r}_1, ..., \vec{r}_N, t)$, as if the $N$ particles were the components of a *single* physical object: an Ansatz going much beyond de Broglie's intuition of objective 3-dimensional single-particle matter waves, on which both Eqs. (6) and (8) are based. The relevant *time-independent* and *time-dependent* Schrödinger equations (6) and (8) were *heuristically* "extended", respectively, in the form

$$\sum_{k=1..N} \frac{\hbar^2}{2m_k} \nabla_k^2 \, \psi + [E - V(\vec{r}_1, ..., \vec{r}_N)] \, \psi = 0 \qquad (6')$$

and

$$i\hbar \frac{\partial \psi}{\partial t} = - \sum_{k=1..N} \frac{\hbar^2}{2m_k} \nabla_k^2 \, \psi + V(\vec{r}_1, ..., \vec{r}_N) \, \psi \,, \quad (8')$$

where $E$ is the total energy of the particle system, and the potential energy $V(\vec{r}_1, ..., \vec{r}_N)$ keeps both external fields and internal interactions into account. Eq. (6') is seen to agree with Eq. (8') *if, and only if,* we "extend" to $\psi(\vec{r}_1, ..., \vec{r}_N, t)$ the same expression (3) which was originally conceived by de Broglie for his *single particle* matter waves: $\psi = u(\vec{r}, E) \, e^{-i E t / \hbar}$, so that

$$i\hbar \frac{\partial \psi}{\partial t} = E \psi \,. \qquad (11)$$

## IV. BOHMIAM MECHANICS

The emergence of the SQM tenets was *accompanied by* de Broglie's interpretation presented in his doctoral Thesis [3] and by Madelung's hydrodynamic alternative [12], and *followed* by Bohm's theory [13-18] (stemming from a de Broglie's suggestion [3]), by de Broglie's return with his "double-solution" proposal [19-21] and by Takabayasi's stochastic approach [22]. The most successful developments were connected with the Bohmian theory [13-18], kept alive for many years by Holland's book [23], and were mainly due to the applicative requirements of the physical-chemistry community [1, 24-32]. In Bohm's theory, a replacement of the form

$$\psi(\vec{r}, t) = R(\vec{r}, t) \, e^{\, i G(\vec{r}, t) / \hbar} \,, \qquad (12)$$

with real $R(\vec{r}, t)$ and $G(\vec{r}, t)$, is performed into the *time-dependent* Schrödinger Eq. (8), assuming $R^2$ to represent

(in the attempt to deviate as little as possible from the Copenhagen orthodoxy) *"the probability density for particles belonging to a statistical ensemble"* [13]. The replacement (12) leads to a *fluid-like* equation system (which we shall omit here for brevity sake) coupled by a *time-dependent* "Quantum Potential" term of the form

$$Q_B(\vec{r}, t) = -\frac{\hbar^2}{2m} \frac{\nabla^2 R(\vec{r}, t)}{R(\vec{r}, t)} \,, \qquad (13)$$

depending on the entire set of eigen-fuctions required by the Born Wave Function $\psi(\vec{r}, t)$. The replacement (12) - shaped on Eq. (3), i.e. on de Broglie's *mono-energetic* and experimentally tested matter waves - aims at dressing the Born Rule with plausibility by depicting $\psi(\vec{r}, t)$ as an individual and objective physical wave, hopefully sharing and generalizing the same experimental evidence of de Broglie's pilot waves (3), although it is not even the solution of an ordinary-looking wave equation. According to Ref. [27], *"Born had an absolutely correct (...) intuition about the meaning of the Wave Function, namely that it guides the particles and it determines the probability of particle positions (...). Born is close to Bohmian mechanics"*.

Being the computation of $Q_B(\vec{r}, t)$ a very hard matter (it was built, for instance, by means of the iterative solution of an infinite set of complex equations [26]) it is often bypassed, in modern Bohmian Mechanics [1, 27], by assuming an equivalent, but more tractable, "*guidance equation*" of the form

$$\frac{d\vec{r}(t)}{dt} = \frac{\vec{V}G(\vec{r}, t)}{m} \equiv \frac{\hbar}{mi} \, Im \left( \frac{\vec{\nabla} \psi}{\psi} \right)$$
$$\equiv \frac{\hbar}{2mi} \frac{\psi^* \vec{\nabla} \psi - \psi \vec{\nabla} \psi^*}{\psi \psi^*} \qquad (14)$$

where $\psi \psi^* \equiv |\psi|^2 \equiv R^2$, and the analytic expression of $\vec{\nabla} G(\vec{r}, t)$ is directly obtained from Eq. (12). The time-integration of Eq. (14) is performed by means of the feedback input, step by step, of the function $\psi(\vec{r}, t)$ obtained from the simultaneous solution of the relevant *time-dependent* Schrödinger equation (8). Recalling that the quantity $\vec{J} \equiv \frac{\hbar}{2mi} (\psi^* \vec{\nabla} \psi - \psi \vec{\nabla} \psi^*)$ represents, in terms of $\psi(\vec{r}, t)$, a probability current density [7, 8], the "guidance velocity" $\frac{d \, \vec{r}(t)}{dt} \equiv \vec{J} / R^2$ turns out to be directed along *"the flux lines along which the probability density is transported"* [27]. The resulting $\vec{r}(t)$ is interpreted however as representing the *exact quantum trajectory* of a single particle, piloted *(à la* de Broglie) by the Born Wave

Function $\psi(\vec{r},t)$, which is interpreted, in its turn, as an *objective physical wave*. It is symptomatic, to be sure, that no objection was ever raised about the consistency of these "quantum trajectories" with the uncertainty principle. Ref. [32] is one of the few Bohmian works admitting that while "*it is impossible to accurately determine the true path pursued by a quantum particle*", the hydrodynamic streamlines provide, at least, a non-disturbing (i.e. "uncertainty respecting") tool to understand their topology. An alternative Bohmian "route", started in Refs. [24-25], directly interprets Bohm's equation system, indeed, as the hydrodynamical description of an objective "probability fluid", and its streamlines as the "quantum trajectories" of a discretized set of small fluid particles, somewhat playing the role of wave-packets in SQM.

In the case of a system of $N$ particles, Bohm's theory makes use of a set of $N$ guidance equations of the form (14), *non-locally* coupled, thanks to Eq. (8') by the Wave Function of the whole system, depending on all the $N$ particles at the same time. In Bohm's words [18], "*the guidance conditions and the Quantum Potential depend on the state of the whole system in a way that cannot be expressed as a preassigned interaction between its parts. As a result there can arise a new feature of objective wholeness. This (...) follows from the fact that the entire system of particles is organized by a common "pool of active information" which does not belong to the set of particles but which, from the very outset, belongs to the whole*".

## V. THE "CLASSICAL" CONNECTION: THE WAVE POTENTIAL "ROUTE"

An approach [33-37] centered on a *point-particle* model (and avoiding therefore the conceptual difficulties of a *wave-packet* representation) has recently stressed that a full exploitation of the *time-independent* Schrödinger equation could provide a straightforward wave-dynamical extension of classical Mechanics. The starting point was the observation that any wave described by a Helmholtz-like equation may be treated in terms of a Hamiltonian set of *exact* ray-trajectories *(bypassing any quasi-optical approximation)* mutually coupled by a monochromatic, *dispersive* function (called "Wave Potential"), encoded in the structure itself of the Helmholtz equation and *acting normally to the ray-trajectories*. The Helmholtz-like structure of the *time-independent* Schrödinger's equation suggests therefore to apply the same method to the determination of the *exact*, *trajectory-based single-particle* dynamics, ruled by a suitable mono-energetic Wave Potential. The fact of acting *normally* to the relevant particle trajectories (a property of which the Bohmian Quantum Potential (13), because of its composite structure, cannot enjoy) allows to view diffraction and interference as energy-preserving exchanges between the longitudinal and transversal components of the particle momentum. The

*exact* point-particle dynamics allowed by the *time-independent* Schrödinger may be accompanied by a *statistical* treatment based on the coefficients $|c_n|^2$ of the solution (10) of the *time-dependent* equation, more or less like Classical Statistical Mechanics is based on Classical Dynamics.

By replacing (3) into (6) and separating real and imaginary parts, the *time-independent* Schrödinger equation (6) may be shown, in fact, to be structurally associated with a *self-contained* Hamiltonian set of *exact* single-particle trajectory equations of the form

$$\begin{cases} \dfrac{d\,\vec{r}}{d\,t} = \dfrac{\partial H}{\partial\,\vec{p}} \equiv \dfrac{\vec{p}}{m} & (15) \\[2mm] \dfrac{d\,\vec{p}}{d\,t} = -\dfrac{\partial H}{\partial\,\vec{r}} = -\vec{\nabla}\,[V(\vec{r})+Q(\vec{r},E)] & (16) \\[2mm] \vec{\nabla}\cdot(R^2\,\vec{p}\,)=0 & (17) \\[2mm] p(t=0)=\sqrt{2\,m\,E} \equiv 2\pi\hbar/\lambda_0 & (18) \end{cases}$$

where *no simultaneous solution of the time-dependent Schrödinger equation is required*, and

$$H(\vec{r},\vec{p},E) \equiv \dfrac{p^2}{2m}+V(\vec{r})+Q(\vec{r},E)=E \quad (19)$$

$$Q(\vec{r},E)=-\dfrac{\hbar^2}{2m}\dfrac{\nabla^2 R(\vec{r},E)}{R(\vec{r},E)}. \quad (20)$$

The *time-independent, energy-dependent* function $Q(\vec{r},E)$, which we call "Wave Potential", turns out to couple together all the relevant particle trajectories, and it may be shown, as a consequence of Eq. (17), that the wave amplitude $R(\vec{r},E)$ and its functions are distributed over the relevant wave-fronts, so that $\vec{p}\cdot\vec{\nabla}\,Q(\vec{r},E)=0$. The Wave Potential $Q(\vec{r},E)$ doesn't cause therefore any wave-particle energy exchange: a property of which the Bohmian *time-dependent* "Quantum Potential" $Q_B(\vec{r},t)$ (13), involving the full set of eigen-energies and eigen-functions, cannot enjoy, in spite of the *formal* coincidence between Eqs. (13) and (20). The two "Potentials" refer indeed to different (point-like and wave-packet, respectively) particle representations.

In *apparent* violation of the uncertainty principle (but, *in effect*, as a simple result [37] of having avoided any wave-packet particle model) the dynamical system (15)-(18) may be time-integrated by assigning the launching values $\vec{r}(E,t=0)$ and $\vec{p}(E,t=0)$ of the particle positions and momenta, together with the wave amplitude distribution $R(\vec{r},E,t=0)$ over a launching surface. The numerical time-integration provides the evolution, step by step, of $\vec{r}(E,t)$ and $\vec{p}(E,t)$, i.e. a full description of the point-particle motion along an exact stationary set of trajectories coupled by the Wave Potential, in the frame of a wave-like stationary phenomenon (pervading in principle the entire physical space) where the omission of the Wave Potential would reduce the Hamiltonian system (15)-(18) to the eikonal approximation [7] of matter waves, i.e. to Classical Dynamics.

A number of examples of exact single particle trajectories obtained *in complete agreement with Schrödinger's equations* are given now by means of the numerical solution of the Hamiltonian equations (15)-(18), performed by assuming, for simplicity sake, a geometry allowing to limit the computation to the *(x, z)*-plane, where both $R(\vec{r},E)$ and its functions satisfy over any wave-front, thanks to Eq. (17), the relation $\partial/\partial z = -\left(p_x/p_z\right)\partial/\partial x$. By expressing the space coordinates $x$ and $z$ in terms of the half-width $w_0$ of the starting slit, Fig. 1 presents, to begin with, the diffraction of a Gaussian matter wave beam launched along the z-axis with $p_x(t=0)=0$; $p_z(t=0)=\sqrt{2mE}$, in the absence of external fields (i.e. for $V(x,z)=0$), in the form $R(x;z=0) \propto exp(-x^2/w_0^2)$, starting from a vertical slit (with half-width $w_0 > \lambda_0$) centered at $x=z=0$. In order to fix ideas, we refer to a case of cold neutron diffraction with $\lambda_0 = 19 \times 10^{-4}\ \mu m$, $w_0 = 11.5\ \mu m$, $\varepsilon \equiv \lambda_0/w_0 \cong 1.65 \times 10^{-4}$.

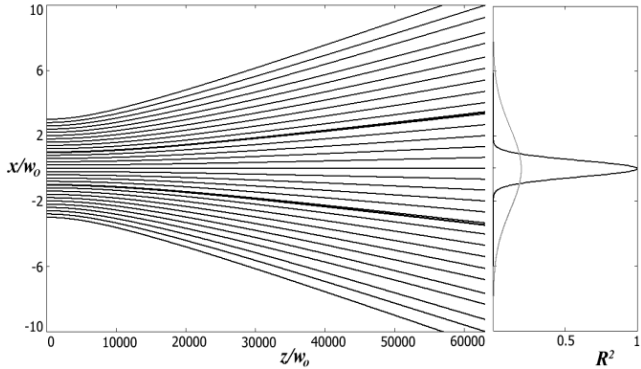

FIG.1.  Diffraction of a Gaussian matter wave beam.

We plotted on the right-hand side the initial and final transverse intensity distributions of the beam, and on the left-hand side the relevant ray-trajectory pattern. The diffractive process due to the Wave Potential consists of the beam gradual widening, while preserving the total kinetic energy.

The two heavy lines represent the trajectories starting (at $z=0$) from the so-called "waist" positions $x/w_0 = \pm 1$, whose numerical values turn out to be in excellent agreement with their well-known *paraxial* analytical expression

$$\frac{x}{w_0} = \pm\sqrt{1+\left(\frac{\lambda_0\, z}{\pi\, w_0^2}\right)^2}\quad.\qquad(21)$$

Fig. 2 refers to the diffraction/interference case of two neighbouring Gaussian coherent wave beams of the form $R(x;z=0) \propto exp[-(\frac{x}{w_0}\pm 1.4)^2]$. We plotted on the right-hand side the initial and final transverse intensity distributions of the beams, and on the left-hand side the relevant ray-trajectory pattern.

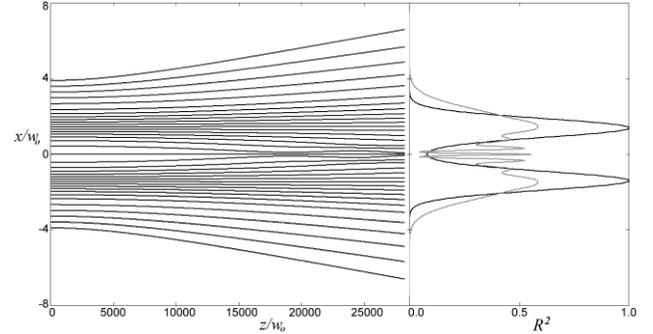

FIG.2.  The case of two neighbouring Gaussians coherent beams.

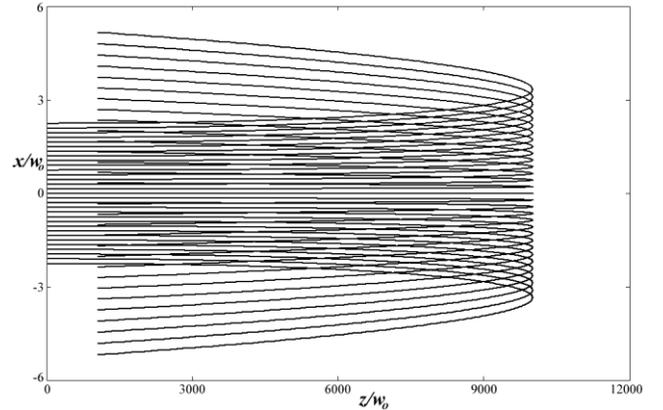

FIG.3.  Gaussian beam launched against a constant field $-F_{oz}$.

Fig. 3 shows, in its turn, the launch, stopping and "backward fall" of the same beam of Fig. 1, traveling in an external potential field of the form $V \equiv V(z) = F_{oz}\, z$, i.e. under a constant force field $-F_{oz}$ acting in the negative *z*-direction.

Starting from *z*=0, the beam travels, for a while, in the positive *z* direction; when $z \cong E/F_{oz}$ it's stopped by the force field, and "falls back" (while continuing its diffractive widening due to the Wave Potential) towards the starting position.

Referring now to:

1) a stationary potential barrier of the *Gaussian* form

$$V = V(z) = V_0 \, exp \, [-2 \, (z - z_G)^2 / d^2 \, ] \; , \qquad (22)$$

(where $z_G = 10000 \, w_0$ is the position of the peak, and $d = 5000 \, w_0$ is the distance between the flexes), and to

2) a *logistic* (step-like) stationary potential function of the form

$$V = V(z) = V_0 \, \left\{ 1 + exp \, [- \, \alpha \, \frac{z - z_L}{w_0} \, ] \right\}^{-1} , \qquad (23)$$

where the parameters $\alpha = 0.002$ and $z_L / w_0 = 10000$ determine, respectively, the slope and the flex position of the continuous line connecting the two asymptotic levels where $V(z \to -\infty) = 0$ and $V(z \to \infty) = V_0$, we plot in Fig. 4 the respective ratios $V(z)/V_0$, and "launch" (from the left hand side) the same beam of Figs. 1 and 3, with total energy $E$, into these external fields.

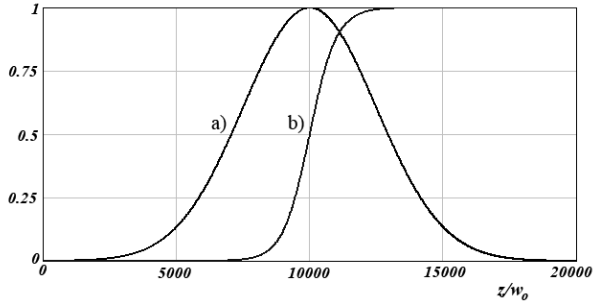

FIG.4. **(a)** Gaussian and **(b)** step-like ratios $V(z)/V_0$ .

In the case (Figs. 5-7) of the *potential barrier* (22), the beam gradually widens under the action of the Wave Potential, and is stopped and thrown back, at a $z$-position lower than $z_G$ where $E = V(z) < V_0$. We omit, for brevity sake, the relevant trajectory plot, because of its similarity with Fig. 3. The most interesting plots are obtained for $E/V_0 \cong 1$. Both when the beam is stopped and thrown back, just before $z = z_G$, for a value of $E/V_0$ *just below 1* (Fig. 5), and when the beam overcomes the potential barrier for a value of $E/V_0$ *just above 1* (Fig. 6), the beam particles spend a part of their time in a narrow fringe region close to the position $z = z_G$, where both the external force $F_z(z)$ and $p_z$ are very close to zero. In these conditions the dominant role is played by the Wave Potential, causing a strong transverse widening of the beam, which is progressively accelerated for $z > z_G$. We finally show in Fig. 7 the case $E/V_0 >> 1$, where the beam overcomes the top of the barrier and undergoes a strong acceleration beyond it.

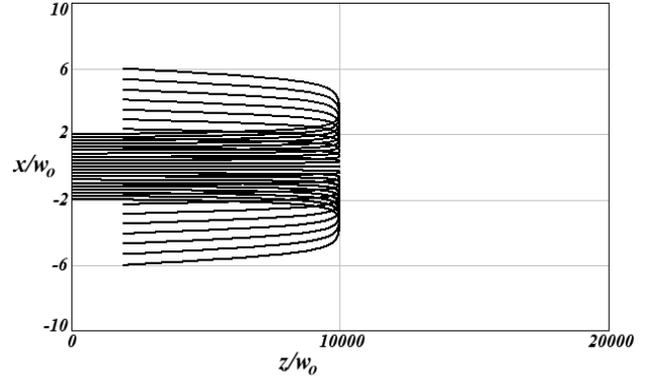

FIG.5. Potential barrier: case $E/V_0$ just below *1*.

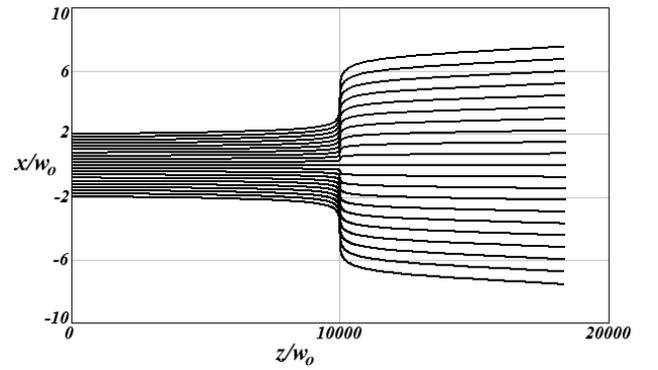

FIG.6. Potential barrier: case $E/V_0$ just above *1*.

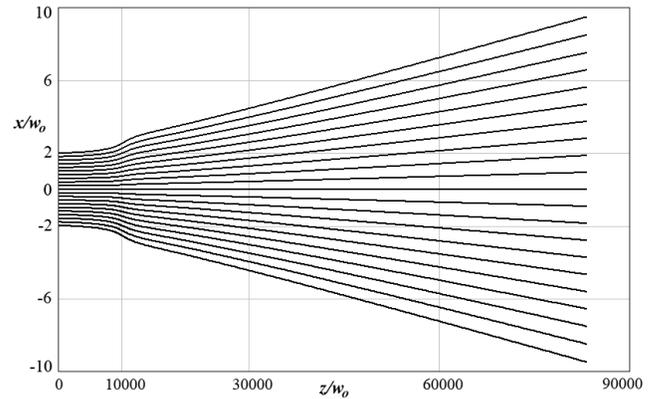

FIG.7. Potential barrier: case $E/V_0 >> 1$ .

In the case (Figs. 8-10) of the *step-like potential* (23), the discussion is quite similar to the one performed for the *potential barrier*, presenting however a few peculiar differences.

The beam gradually widens under the action of the Wave Potential, and is stopped and thrown back, for $E = V(z) < V_0$ with a behavior (quite analogous to the one of Fig. 3) whose plot we omit here, once more, for brevity sake. Once again, the most interesting plots are obtained for $E / V_o \cong 1$.

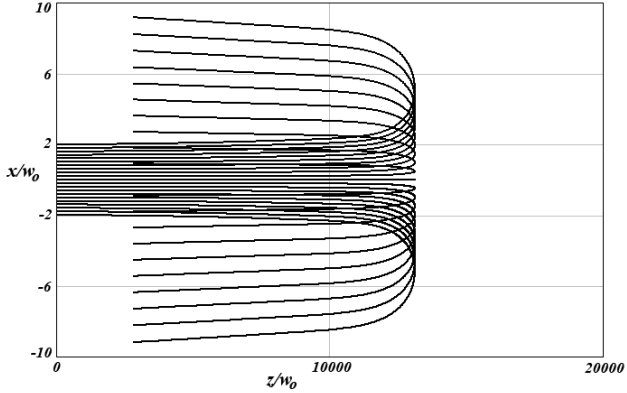

FIG.8.  Step-like potential: case $E / V_0$ *just below 1*.

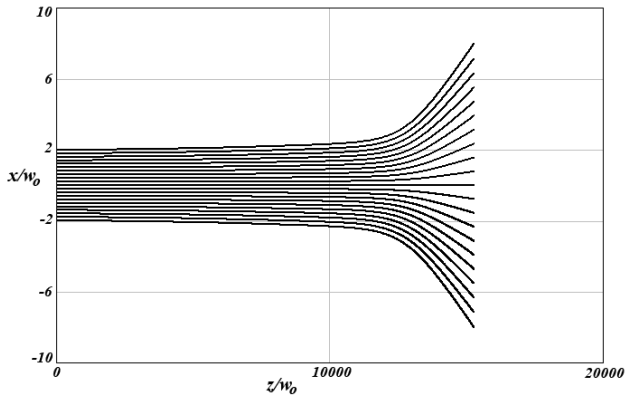

FIG.9.  Step-like potential: case $E / V_0$ *just above 1*.

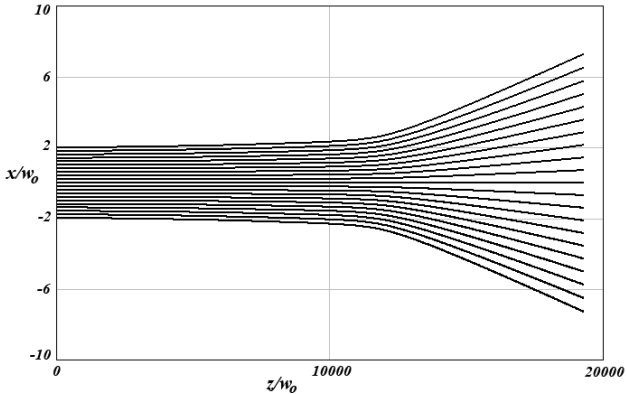

FIG.10.  Step-like potential: case $E / V_0 >> 1$.

Both in Fig. 8 (beam stopped and thrown back for a value of $E / V_0$ *just below 1*) and Fig. 9 (beam overcoming the potential step for a value of $E / V_0$ *just above 1*) the beam particles spend a part of their time in a narrow fringe region around a position (close to the top of the step) where both the external force $F_z(z)$ and $p_z$ are very close to zero, and the dominant role is played, once more, by the Wave Potential, causing a strong transverse widening of the beam. The main differences from the previous case stand in the fact that while particles reaching the top of the potential barrier with $p_z \cong 0$ receive, from there on, a positive forward push, particles getting the top of the step function with $p_z \cong 0$ are (and remain) endowed with a basically transverse momentum. The beam doesn't meet a further force field, and widens under the action of the Wave Potential alone: a behavior which goes on, in Fig. 10, for $E / V_0 >> 1$.

Let us finally come to the case of particles moving in an external stationary potential field $V(x, z)$ representing a *lens-like* focalizing structure. We previously recall [8, 9, 37] that, by simply performing the replacements

$$\frac{2 m E}{\hbar^2} \to \frac{p_0^2}{\hbar^2} \to k_0^2 \; ; \; 1 - \frac{V(\vec{r})}{E} \to n(\vec{r})^2 \; , \qquad (24)$$

the *time-independent* Schrödinger equation (6) takes on the form of the Helmholtz equation

$$\nabla^2 u(\vec{r}) + [n(\vec{r}) k_0]^2 \, u(\vec{r}) = 0 \qquad (25)$$

holding for electromagnetic waves with $k_0 = 2\pi / \lambda_0$ in a medium with refractive index $n(\vec{r})$, while the respective *eikonal limits* transform according to the correspondence

$$p^2 \cong 2 m E (1 - V / E) \leftrightarrow k^2 \cong k_0^2 \; n^2 \; . \qquad (26)$$

We assign therefore a refractive index of the form [38]

$$n(x, z) = 1 + exp \left[ - \left( \frac{x}{L_x} \right)^2 - \left( \frac{z - Z_0}{L_z} \right)^2 \right] \qquad (27)$$

and assume

$$V(x, z) = E \, [1 - n(x, z)^2] \qquad (28)$$

in Eq. (16).

We present in Fig. 11 and Fig. 12 the numerical results obtained (with a suitable choice of the parameters $L_x$ , $L_z$ and $Z_0$) for the same particle beam of Fig. 1 *by neglecting and by taking into account, respectively,* the Wave Potential term $Q(\vec{r}, E)$, whose diffractive effect is seen to replace the point-like eikonal focus by a finite focal waist. Fig. 13 shows, in its turn, the progressive intensity sharpening of the focused beam.

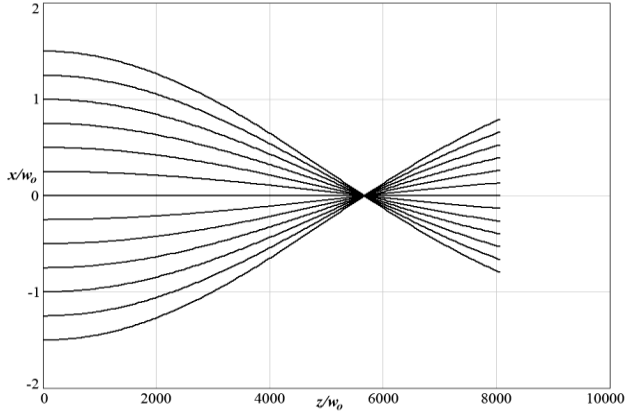

FIG.11. Lens-like potential: eikonal (point-like) focusing of a Gaussian matter wave beam.

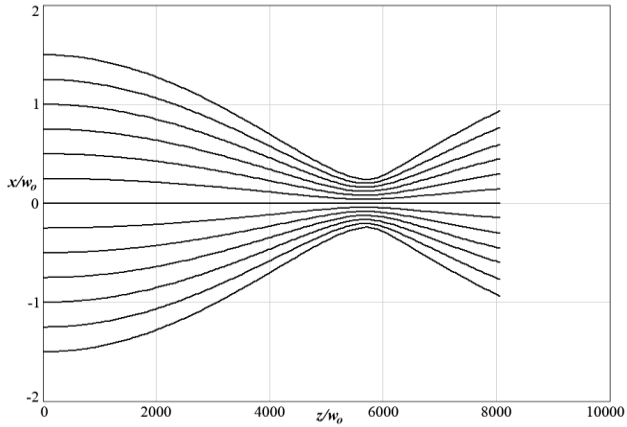

FIG.12. Lens-like potential: finite waist focusing of a Gaussian matter wave beam.

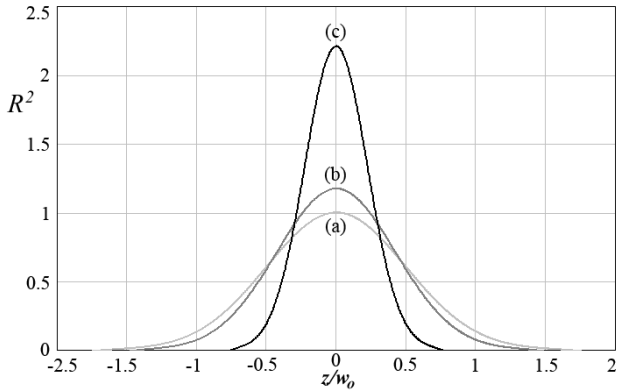

FIG.13. Lens-like potential: progressive intensity sharpening of a focused Gaussian matter wave beam.

## VI. BEYOND QUASI-OPTICS

The *exact* trajectory-based solutions of the Hamiltonian system (15)-(18), presented in the previous Section for the dynamics of *point-like particles* piloted by de Broglie's *monochromatic matter waves*, are analogous to the ones (concerning *monochromatic electromagnetic waves*) obtained at the Institute of Plasma Physics of the C.N.R. of Milan [38-42] by one of the Authors (AO), within the limits of a complex-eikonal *quasi-optical approximation* originally proposed in Refs. [43-44] and successfully extended to the propagation of *gyro-resonant e.m. waves* launched into magnetoactive *thermonuclear plasmas* for diagnostic and/or plasma-heating purposes. A toroidal (3D) ray-tracing code provided a satisfactory description of the finite-waist formation and diffractive self-widening processes affecting the transmission, reflection and absorption of high frequency *electromagnetic Gaussian beams*, in experiments of crucial interest for the beam directivity control and for the stabilization of potentially disruptive magnetohydrodynamic modes in fusion devices. The *quasi-optical* analysis presented in Refs. [38-42] was also applied, in more recent times [45], to the Doppler backscattering microwave diagnostic system installed on the Tokamak TORE SUPRA of Cadarache, waiting for the completion of the ITER prototype of fusion reactor.

Although a quasi-optical analysis was originally applied to the quantum case in Ref. [46], with a set of results quite similar to the ones of Sect.V of the present paper, any quasi-optical approximation is avoided in the present work by the use of the Wave Potential approach.

## VII. CONCLUSION

Our present approach is characterized by a *mono-energetic* "Wave Potential" function acting normally to the relevant *point-particle* trajectories: a property (allowing to pilot the particle motion without modifying its energy) which is not shared by the Bohmian "Quantum Potential", involving the entire set of eigen-energies of the wave trains composing a *wave-packet*.

---

**TABLE I. Bohmian (wave-packet) trajectories**

$$\frac{d\,\vec{r}}{d\,t} = \frac{\hbar}{m\,i}\,Im\,(\frac{\vec{\nabla}\,\psi}{\psi})$$

$$i\,\hbar\,\frac{\partial\,\psi}{\partial\,t} = -\,\frac{\hbar^2}{2\,m}\,\nabla^2\,\psi + V(\vec{r})\,\psi$$

---

**TABLE II. "Exact" (point-particle) trajectories**

$$\frac{d\,\vec{r}}{d\,t} = \frac{\vec{p}}{m}$$

$$\frac{d\,\vec{p}}{d\,t} = -\,\vec{\nabla}[\,V(\vec{r}) - \frac{\hbar^2}{2\,m}\,\frac{\nabla^2 R(\vec{r},E)}{R(\vec{r},E)}\,]$$

$$\vec{\nabla}\cdot(R^2\,\vec{p}) = 0$$

---

We summarize and compare the Bohmian and our own approach in Tables I and II, respectively, holding for particles moving in an external stationary potential field $V(\vec{r})$. It is seen, in conclusion, that:

1. *the Bohmian approach* provides, by means of its "guiding equation", a set of *probability flow-lines* resulting from the entire ensemble of eigen-functions composing a *wave-packet*, and built up by the simultaneous solution of Schrödinger's *time-dependent* equation, while

2. *our own approach* provides (by means of a set of ordinary-looking dynamic equations encoded in Schrödinger's *time-independent* equation) the *exact trajectories* of *point-particles* with assigned energy *E*, guided by the relevant (monochromatic) de Broglie's wave.

Let us also remind that the *exact, point-particle, trajectory-based* Hamiltonian equations associated with the relativistic time-independent *Klein-Gordon* equation (and reducing, of course, to eqs. (15)-(18) in the non-relativistic limit) were obtained (by the Authors of the present paper) in Ref. [36].

Besides allowing an *exact* forward step with respect to the *quasi-optical approximation* employed in the treatment of *classical* waves, we provide, in conclusion, a consistent *wave-mechanical* extension of *point-particle* Classical Dynamics avoiding any *wave-packet* representation: a representation, indeed, about which Born himself [47] wrote that "*it tempts us to try to interpret a particle of matter as a wave-packet due to the superposition of a number of wave trains. This tentative interpretation, however, comes up against insurmountable difficulties, since a wave-packet of this kind is in general very soon dissipated*".

---